\documentclass[twocolumn,showpacs,preprintnumbers,amsmath,amssymb,prb,a4]{revtex4}

\usepackage{graphicx}

\newcommand{\Na}{NaV$_2$O$_5$}
\newcommand{\Ca}{CaV$_2$O$_5$}
\newcommand{\Mg}{MgV$_2$O$_5$}

\begin{document}
\title{Optical properties and Raman scattering of vanadium ladder compounds}
\author{J.~Spitaler}
\email{juergen.spitaler@uni-graz.at}
\author{E. Ya. Sherman}
\author{C. Ambrosch-Draxl}
\affiliation{Institut f\"{u}r Theoretische Physik, University Graz, Universit\"{a}tsplatz
5, A-8010 Graz, Austria}
\author{H. - G. Evertz}
\affiliation{Institut f\"{u}r Theoretische Physik, Technical University Graz, Petersgasse 16, A-8010 Graz, Austria}
\date{\today}
\begin{abstract}
We investigate electronic and optical properties of  the V-based ladder compounds \Na, the iso-structural \Ca\, as well as \Mg, which differs from  \Na\ and \Ca\
 in the $c$ axis stacking. We calculate {\em ab initio} the A$_{g}$ phonon modes in these compounds as a basis for the investigation of the electron-phonon and spin-phonon coupling. The phonon modes together with the dielectric tensors as a function of the corresponding ion displacements are the starting point for the calculation of  the A$_g$ Raman scattering. 
\end{abstract}

\maketitle

\section{Introduction}\label{intro}

The vanadium ladder structures \Na, \Ca, and \Mg\ are low-dimensional compounds with unusual physical properties due to a strong coupling of charge, spin, and lattice degrees of freedom.   \Na\ and \Ca\ are iso-structural (space group $D_{2h}^{13}$  $(Pmmn)$) while \Mg\ has the symmetry group $Cmcm$.The main structural element of all these compounds is a simple ladder containing V-O rungs and V-O legs providing a possibility for electrons to hop within a rung and between the rungs of one ladder, whereas the hopping matrix element between different ladders is considerably smaller. The highest occupied states are formed by $d_{xy}$ electrons of V mixed with oxygen $2p$ orbitals. 
 \Ca\ and \Mg\ are both  half-filled compounds with similar electronic properties, i.e. one $d_{xy}$ electron per V site. However, they show different magnetic behavior \cite{korotin00}.  	
The quarter-filled \Na\, where one $d_{xy}$ is shared by two V sites within a rung \cite{smolinski98},  undergoes a spin-Peierls transition at $T_c=35$ K \ \cite{isobe96}  accompanied by a charge ordering and a large lattice distortion. In this compound, phonons induce strong charge fluctuations near $T_c$ that lead to fluctuations in the spin-spin exchange $J$ \cite{Sherman99}.  Therefore, investigation of lattice dynamics and electron (spin)\ -\ phonon coupling in ladder compounds can provide a clue to their properties. 

Investigations of these compounds by optical spectroscopies - ellipsometry and Raman scattering - have shown many interesting features. The former gives the dielectric function, whereas the latter provides information about phonons and low-energy excitations and their coupling to light. For \Na\ the phonon Raman spectra demonstrate very large changes below the transition point \cite{fischer99} such as variations in the electronic background \cite{Sherman01} and  the appearance of intense new Raman active modes indicating a strong coupling of the lattice to the order parameter of the transition \cite{fischer99,Sherman99}. With the aim to understand similarities and differences in the optical properties of \Na,  \Ca, and \Mg\ we perform {\it ab initio} investigations within density functional theory (DFT). We calculate the fully symmetric A$_{g}$ phonon modes and their influence on the electronic structure and the dielectric function. Such phonon-induced modulations of  the polarizability are a measure for electron-phonon coupling and the source of phonon Raman scattering of light.  
Although the vanadates are considered to be correlated materials\cite{mazurenko02}, the results obtained  in the framework of DFT show good agreement with experiment in terms of their lattice dynamics and optical properties. Since similar findings have been reported earlier for high temperature superconductors\cite{cad2002} such investigations are justified.

\section{Computational details}
All calculations are performed within DFT using the full-potential aug\-men\-ted plane waves + local orbitals (FP-APW+lo) \cite{sjostedt00} formalism implemented in the WIEN2k code \cite{wien2k}.  Exchange and correlation terms are described within the generalized gradient approximation (GGA) \cite{perdew96}. All atomic positions of the three structures have been relaxed as a starting point for the investigation of the phonon modes within the frozen-phonon approach. 
We have displaced the atomic coordinates from their equilibrium positions, where for each degree of freedom two displacements in both, positive and negative direction, respectively, have been used. The corresponding forces established the dynamical matrix for the A$_{g}$ phonons.
The dielectric tensors are computed within the random phase approximation with the Kohn-Sham eigenstates taken as an approximation to the single particle states.

\section{Dielectric function}

\begin{figure}[htb]
\begin{center}
\includegraphics[width=0.45\textwidth]{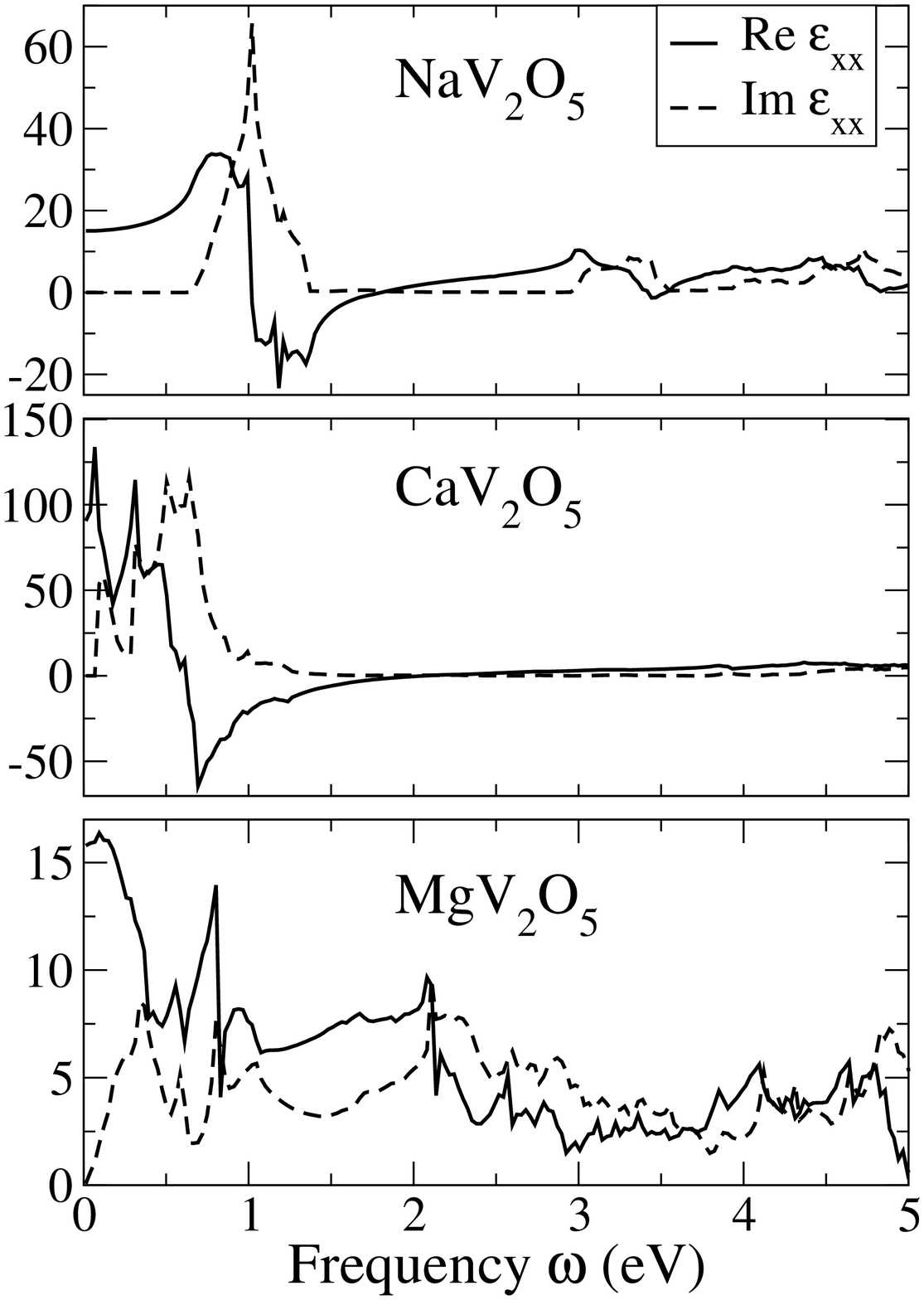}
\end{center}
\caption{ The dielectric function $\varepsilon_{xx}(\omega)$ for \Na, \Ca, and \Mg. The $x$-axis is parallel to the rungs.}
\label{fig:eps_xx}
\end{figure}

\begin{figure}[htb]
\begin{center}
\includegraphics[width=0.45\textwidth]{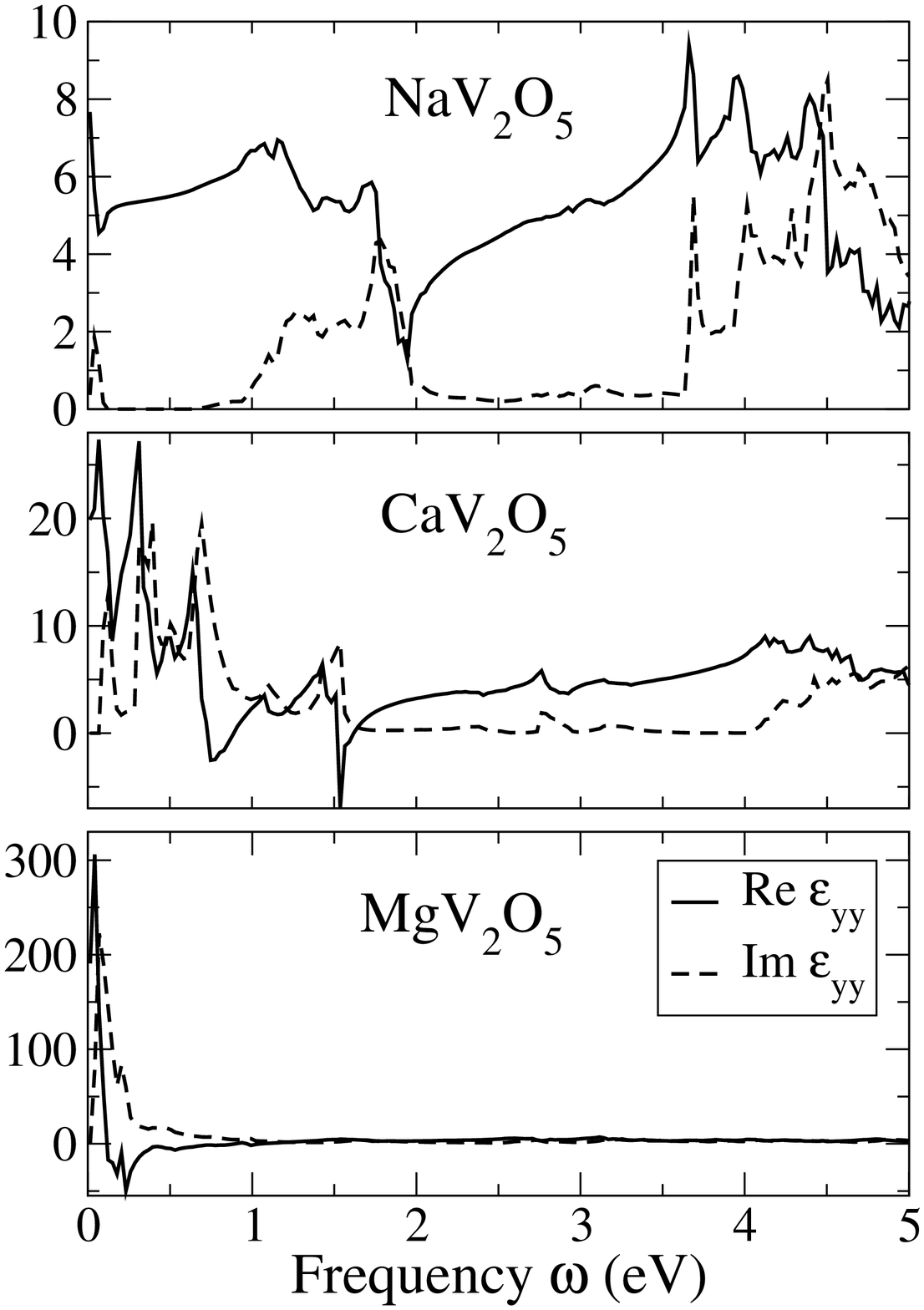}
\end{center}
\caption{ The dielectric function $\varepsilon_{yy}(\omega)$ for \Na, \Ca, and \Mg. The $y$-axis is parallel to the ladders.}
\label{fig:eps_yy}
\end{figure}

Fig.~\ref{fig:eps_xx} and Fig.~\ref{fig:eps_yy} demonstrate the diagonal in-plane components of the 
dielectric function $\varepsilon_{xx}(\omega)$ and $\varepsilon_{yy}(\omega)$ of the three compounds investigated. The real part of $\varepsilon(\omega)$ was obtained by the Kramers - Kronig relation from its imaginary part, where no broadening of the optical transitions was assumed. The strong absorption peak in the imaginary part of $\varepsilon_{xx}(\omega)$ at $\omega$ close to 1 eV and another weak feature in Im $\varepsilon_{yy}(\omega)$ at $\omega$ $\approx$ 1.25 eV
are due to transitions between the bands originating from bonding and antibonding states within one rung separated by $\approx$ 2$|t_{\perp}|$ in energy, where $t_{\perp}$ is the in-rung hopping matrix element. Due to the dispersion of the bands the peak position is larger than 2$|t_{\perp}|$. Our results correspond to $t_{\perp}=-0.35$ eV, in agreement with Ref. \onlinecite{smolinski98}. At the same time, a peak at 1.75 eV in ${\rm Im}$ $\varepsilon_{yy}(\omega)$ arises from transitions within the legs. For \Na\  the calculated dielectric functions are in good agreement with available experimental data \cite{Presura00}. \Ca\ demonstrates similar features, while for \Mg\ the interpretation is more complicated. 

\section {A$_g$ phonon modes}

The A$_g$ phonon modes preserve the symmetry of the unit cell and in particular the inversion symmetry,
and, therefore, are seen in the Raman spectra. Table \ref{tab:phonons} presents the calculated and measured eigenfrequencies of these modes for all three compounds.  In all modes both, V and O displacements are considerably involved in the eigenvectors. The apical oxygen ions only contribute to the highest-frequency modes ($\omega_{\rm ph}>900$ cm$^{-1}$), thereby vibrating along the $z$-axis.
The theoretical eigenfrequencies of \Na\ and  \Ca\ for all but the lowest mode are in very good agreement with experiment, while the agreement for the \Mg\ modes is worse. The eigenvectors of all three compounds correspond well to the assignment given in Ref. \onlinecite{popovic99-ssc}. 
   	 
\begin{table*}[hbt]
\begin{tabular}{|cc|cc|cc|c|}
\hline
\multicolumn{2}{|c|}{ \Na } & \multicolumn{2}{|c|} {\Ca}  & \multicolumn{2}{|c|} {\Mg}  &\\ 
\hline
\multicolumn{2}{|c|}{$\omega_{\rm ph}$ (cm$^{-1}$)}  & \multicolumn{2}{|c|}{$\omega_{\rm ph}$ (cm$^{-1}$)}  & \multicolumn{2}{|c|}{$\omega_{\rm ph}$ (cm$^{-1}$)}  &assignment \cite{popovic99-ssc} \\ 
Exp.\cite{fischer99,popovic02}  & Theory &Exp.\cite{popovic02} & Theory &Exp.\cite{popovic02}& Theory &\\
 \hline
970, 969& 996 & 932 & 890 & 1002 & 702  & V-O3 stretching \\
530, 534& 514 & 539 & 516 &  536  &  566 & V-O2 stretching \\
450, 448& 464 & 470 & 439 &  478  &  400 & V-O1-V bending \\
422, 423& 415 & 422 & 413 &   414 &  375 & O-V-O bending\\
304, 304& 313 &    ?   & 308 &   ?     &  330 & O-V-O bending\\
230, 233& 230 & 236 & 268 &  233 &   310 & O-V-O bending\\
178, 179& 171 & 139 & 197 &   -     &   209 & Me $\parallel$ c\\
90, 90 & 130  &   90  & 140 &  98   & 116  & chain rot.\\[2mm]

\hline
\end{tabular}
\caption{Experimental and calculated phonon frequencies $\omega_{\rm ph}$ for all three ladder compounds and the assignment of the modes.}
\label{tab:phonons} 
\end{table*}

\section {Phonon-induced changes in the dielectric function}

\begin{figure}[htb]
\vspace{5mm}
\begin{center}
\includegraphics[width=0.45\textwidth]{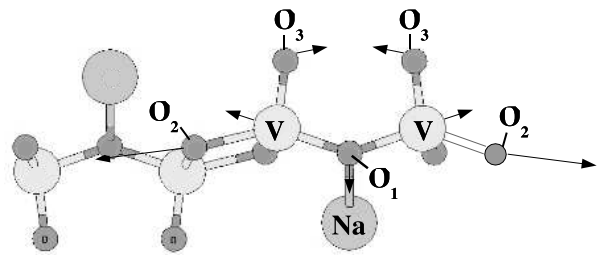}
\end{center}
\caption{Illustration of the in-plane O2-V stretching phonon mode. The lengths of the arrows are proportional to the respective components of the eigenvector.}
\label{fig:mode}
\end{figure}

\begin{figure}[htb]
\vspace{8mm}
\begin{center}
\hspace{-5mm}
\includegraphics[width=0.42\textwidth]{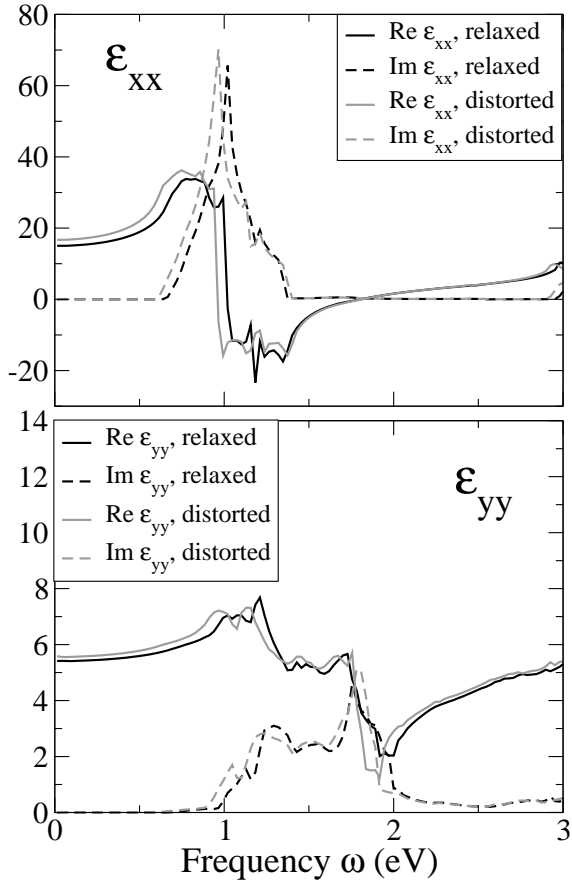}
\end{center}
\caption{Effect of the V-O2 stretching on the dielectric function in \Na.}
\label{fig:na_changes}
\end{figure}

\begin{figure}[htb]
\vspace{8mm}
\begin{center}
\hspace{-5mm}
\includegraphics[width=0.42\textwidth]{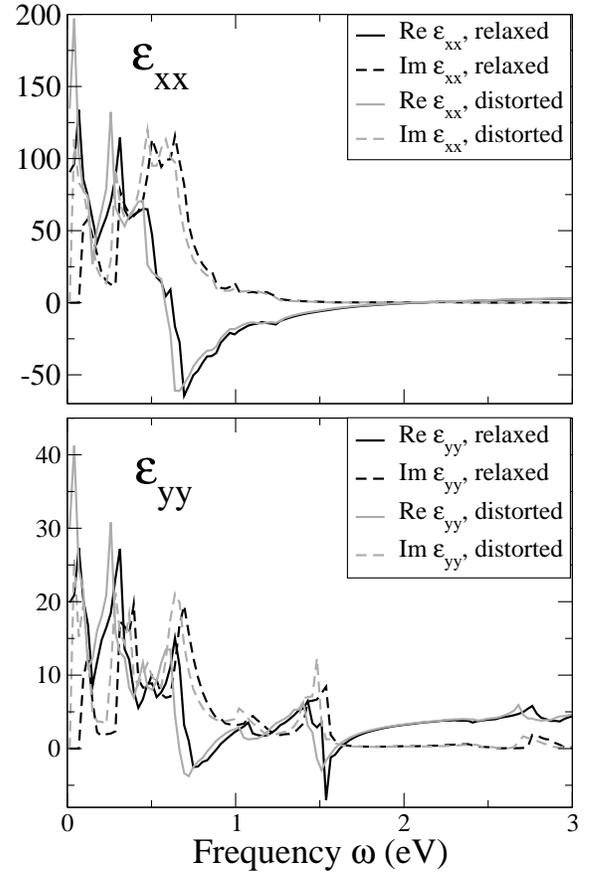}
\end{center}
\caption{Effect of the V-O2 stretching on the dielectric function in \Ca.}
\label{fig:ca_changes}
\end{figure}
Raman scattering arises due to phonon-induced changes in the band structure,
and, correspondingly, in the dielectric function $\delta\varepsilon_{\alpha\alpha}(\omega)$, where $\alpha$ are the Cartesian indices. The intensity of the one-phonon scattering $I_{R}$  for the incident and scattered light parallel to the $\alpha$ axis can be written as 
\begin{equation}
I_{R}\propto 
\left| \langle1|
\frac{\partial \varepsilon_{\alpha\alpha}} {\partial Q} 
\hat{Q}
|0\rangle 
\right|^2.
\end{equation}
Here $Q$ is the coordinate of the phonon mode, and the matrix element represents transitions between the states $|0\rangle$ and $\langle1|$, which are the phononless ground state and the one-phonon state, respectively. 

 Fig.~\ref{fig:mode} represents the ion displacements in the V--O2 stretching mode which is taken as an example to demonstrate the influence of the phonons on the dielectric function. Figs.~\ref{fig:na_changes} and \ref{fig:ca_changes} show the changes in the $xx$ and $yy$ components of the dielectric function of \Na\ and \Ca, respectively. The black lines correspond to the situation of the undisplaced positions, whereas the grey lines reflect  the distorted lattice with the displacement of the O2 ion 
along the $x$-axis taken to be 0.02~\AA, which is of the order of magnitude of the zero-point vibrations. The V ion is shifted correspondingly according to the eigenvector of the mode.
As it is seen in the figure, this mode causes big changes in $\varepsilon_{xx}(\omega)$ at $\omega$ close to the resonance at 1 eV which corresponds to the intra-rung transitions due to a modulation of the $t_{\perp}$ hopping within a rung. 
A general feature of the phonon-induced changes in the dielectric functions is that all the peaks are considerably shifted. This fact implies large values for the first derivatives of the dielectric functions, which govern the Raman scattering intensities. We therefore predict strong resonance effects in phonon Raman scattering close to these peak positions. E.g. the 1~eV region of \Na\ should exhibit resonances  in the $xx$ component. For \Ca\ we expect even more resonances since the peak structure of the dielectric function is more pronounced. Due to our findings we suggest Raman scattering experiments to be performed below 2 eV, the energy region where we expect very high Raman intensities. At the same time, we can explain the experimentally observed intensity enhancement for \Na\ when changing the frequency of the laser beam from 1.95 to 2.4 eV in $xx$ polarization\cite{fischer99}, since the latter frequency is closer to the peak in Re $\varepsilon_{xx}(\omega)$ at $\approx$3 eV as can be seen in the upper panel of Fig.\ref{fig:eps_xx}.

\section {Conclusions}

We have performed {\it ab initio}  calculations for the optical properties and A$_{g}$ phonon modes in the ladder compounds \Na, \Ca, and \Mg. For \Na\ the calculated $\varepsilon(\omega)$ compares well with available experimental data. For \Na\ and \Ca\  the phonon frequencies are in a very good agreement with the experiment, while for \Mg\ the agreement is not as good. Due to the strong influence of the ion displacement on the dielectric function we predict pronounced resonance effects in phonon Raman scattering in the energy range below 2~eV. 

\phantom{ }
\noindent{\bf Acknowledgment}  

We acknowledge support from the Austrian Science Fund (project P15520), and the EU (RTN network EXCITING, contract HPRN-CT-2002-00317). 

\bibliography{literature}

\end{document}